\documentclass[3p,times,procedia]{elsarticle}
\usepackage{nupha_ecrc}
\usepackage{amsmath}

\volume{00}

\firstpage{1}

\journalname{Nuclear Physics A}

\runauth{S. Shi, C. Gale, and S Jeon}

\jid{nupha}

\jnltitlelogo{Nuclear Physics A}

\usepackage{amssymb}

\begin{document}

\begin{frontmatter}



\dochead{XXVIIIth International Conference on Ultrarelativistic Nucleus-Nucleus Collisions\\ (Quark Matter 2019)}
\title{From Chiral Kinetic Theory To Spin Hydrodynamics}


\author{Shuzhe Shi} 
\author{Charles Gale} 
\author{Sangyong Jeon}

\address{Department of Physics, McGill University, 3600 University Street, Montreal, Quebec H3A 2T8, Canada.}

\begin{abstract}
The total angular momentum is conserved in the evolution of the Quark-Gluon Plasma (QGP) created in heavy-ion collision, and consists of two sectors: the orbital angular momentum (OAM) caused by kinetic motion, and the spin, an intrinsic degree of freedom of quarks and gluons. Microscopic scattering processes allow the conversion between these two components, hence the spin density could eventually have non-trivial influence on the QGP evolution. A hydrodynamic theory, with the spin polarization effect properly taken into account, is required to quantitatively study the polarization rate for observed hadrons, e.g. the $\Lambda$-hyperon. In this work, we start with chiral kinetic theory and construct the spin hydrodynamic framework for a chiral spinor system. We obtain the equations of motion of second-order dissipative relativistic fluid dynamics with non-trivial spin polarization density. 
\end{abstract}


\end{frontmatter}

\section{Introduction}
\label{sec.intro}

Relativistic heavy-ion collisions provide a special environment to study the strong interaction. 
In such experiments, a new phase of matter --- the Quark-Gluon Plasma (QGP) --- is created. 
Recently the STAR Collaboration reported measurement of the non-vanishing polarization rate of $\Lambda$-hyperons at the Relativistic Heavy Ion Collider \cite{STAR:2017ckg,Adam:2018ivw}. 
This result reflects an extremely vortical fluid flow structure in the QGP produced in the Au-Au collisions, and has attracted significant interest and generated wide enthusiasm. 
In addition, detailed measurement of the spin polarization, in particularly the longitudinal polarization at different azimuthal angles, disagrees with current theoretical expectation \cite{Adam:2019srw,Becattini:2017gcx,Xia:2018tes}. 

In theoretical attempts (e.g. \cite{Shi:2017wpk,Li:2017slc,Karpenko:2016jyx}) to compute the hadron polarization rate, one typically makes the assumption that hadrons are created according to the thermal equilibrium distribution for particles in a locally rotating fluid, while the viscous corrections induced by off-equilibrium effects are not taken into account. Also, studies assume that the spin degrees of freedom of either hadrons or partons have negligible influence on the dynamical motion of the medium. More theoretical studies with dissipative corrections are required to understand the discrepancy alluded to above, and to describe the vortical structure of QGP. Consequently, we propose to develop a relativistic hydrodynamic theory with spin degrees of freedom.

Being a macroscopic theory based on conservation laws and the second law of thermodynamics, hydrodynamics needs the guidelines of kinetic theory to correctly reflect microscopic processes.
In a massless fermion system, the microscopic transport processes are described by the Chiral Kinetic Theory (CKT)~\cite{Stephanov:2012ki,Liu:2018xip,Huang:2018wdl}.
In this work we first start with CKT and construct the ideal spin hydrodynamic framework for a chiral spinor system, which will be shown in Sec.~\ref{sec.ideal}.
Then in Sec.~\ref{sec.viscous}, we construct the dissipative quantities from the distribution function, and obtain their equation of motion based on conservation rules and chiral kinetic equations.

\section{Ideal Spin Hydrodynamics}\label{sec.ideal}
To describe the collective behavior of spinors taking into account spin degrees of freedom, one can start by defining the Wigner operator:
\begin{equation}
	W_{ab}(x,p) \equiv \left< \int \mathrm{d}^4y\, e^{\frac{i}{\hbar}p\cdot y} \widehat{\bar\psi}_{b}(x+\frac{y}{2})\widehat{\psi}_{a}(x-\frac{y}{2}) \right>,
\end{equation}
and decompose the Wigner function in the Clifford basis,
\begin{equation}
W \equiv 
	\frac{1}{4}\Big( \mathcal{F} + i\,\mathcal{P} \, \gamma^5 + \mathcal{V}_\mu \, \gamma^\mu 
	+  \mathcal{A}_\mu \,\gamma^5 \gamma^\mu + \frac{1}{2} \mathcal{L}_{\mu\nu} \, \Sigma^{\mu\nu} \Big) \,,
\end{equation}
where  $\mathcal{F}$, $\mathcal{P}$, $\mathcal{V}$, $\mathcal{A}$, and $\mathcal{S}$ are known as the Clifford components.
By doing so, the hydrodynamic quantities --- the current, axial current, energy-momentum tensor, and the spin tensor current --- can be expressed respectively as: 
\begin{eqnarray}
\begin{split}
&J^{\mu} \equiv \langle \bar\psi \gamma^\mu \psi \rangle = \int \frac{\mathrm{d}^4p}{(2\pi)^4} \mathcal{V}^\mu \,,&\qquad
&T^{\mu\nu} \equiv \langle \bar\psi (i\gamma^\mu D^\nu) \psi \rangle = \int \frac{\mathrm{d}^4p}{(2\pi)^4} p^\nu \mathcal{V}^\mu \,,\\
&J^{\mu}_{A} \equiv \langle \bar\psi \gamma^\mu\gamma^5 \psi \rangle = \int \frac{\mathrm{d}^4p}{(2\pi)^4} \mathcal{A}^\mu \,,&\qquad
&S^{\lambda\mu\nu} \equiv \frac{1}{4} \langle \bar\psi \{\gamma^\lambda , \Sigma^{\mu\nu} \} \psi \rangle
	= \frac{1}{2}\epsilon^{\lambda\mu\nu\sigma} \int \frac{\mathrm{d}^4p}{(2\pi)^4} \mathcal{A}_\sigma \,. 
\end{split}
\end{eqnarray}
Especially, in a system with Dirac spinors, the spin density current follows once the axial charge density, accounting for the imbalance between right-handed (RH) and left-handed (LH) particles, is defined.

For free Dirac Spinor in the absence of an external field, the equations of motion for Clifford components can be obtained from the Dirac equation.
They couple all Clifford components, and detailed expression can be found in e.g. Ref.~\cite{Huang:2018wdl}.
In that work, the authors employ a semi-classical expansion (i.e. $\hbar$ expansion) in {\it the massless limit} ($m=0$), and derive the Chiral Kinetic Equations up to the leading order in $\hbar$.
In first order CKT, the LH and RH components evolve independently:
\begin{equation}
\mathcal{J}_\pm^\mu
	\equiv \frac{1}{2}(\mathcal{V}^\mu \pm \mathcal{A}^\mu) 
	= 4\pi \delta(p^2) \Big( p^\mu \pm \hbar \frac{\epsilon^{\mu\nu\rho\sigma}p_\rho u_\sigma}{2 p \cdot u} \partial_\nu \Big)\, f_\pm \,,
\end{equation}
where $f_\pm$ are the RH/LH particle densities, defined as the $p^\mu$-proportional section of the corresponding chirality current $\mathcal{J}_\pm^\mu$, and follow the chiral kinetic equations (details see e.g. \cite{Huang:2018wdl,Hidaka:2016yjf}):
\begin{equation} \label{eq.cke}
\Big[p^\mu \partial_\mu \pm
	\hbar \frac{\epsilon^{\mu\nu\rho\sigma} p_\nu (\partial_\rho u_\sigma)}{4\, u \cdot p} \partial_\mu
	\Big] \, f_\pm = 0 \,.
\end{equation}
Particularly, $u$ is a time-like arbitrary auxiliary Lorentz-vector field to construct the $p^\mu$-perpendicular components, and could depend on space-time $x$ in a non-trivial way.
As discussed in detail in Ref.~\cite{Huang:2018wdl}, the explicit form of distribution function $f$ depends on the choice of $u$, which is referred to as the \textit{side-jump} effect \cite{Chen:2015gta}. 
Yet, it could be shown that physical observables like conserved currents are independent of the choice of $u$. 
To better represent the spin polarization effect for particles in the fluid co-moving frame, it's more convenient to take the fluid velocity to be the auxiliary field $u$.

The equilibrium distribution functions has been derived (see e.g.~\cite{Liu:2018xip}), and is
\begin{equation}
f_{\mathrm{eq},\pm}(p) 
	= \frac{1}{\exp[\frac{p\cdot u - \mu_\pm}{T} \pm \hbar \frac{\epsilon^{\mu\nu\rho\sigma}\varpi_{\mu\nu}u_\rho p_\sigma}{4\,u\cdot p}]+1} \,,
\end{equation}
where $\varpi_{\mu\nu} \equiv \frac{1}{2}\Big(\partial_\nu \frac{u_\mu}{T} - \partial_\mu \frac{u_\nu}{T} \Big)$ is the thermal vorticity tensor, $T$ is the temperature, and $\mu^\pm$ are RH/LH chemical potentials, respectively.
We further define the vorticity vector $\omega^\mu \equiv -\frac{T}{2}\epsilon^{\mu\nu\rho\sigma}u_\nu \varpi_{\rho\sigma} = \frac{1}{2}\epsilon^{\mu\nu\rho\sigma}u_\nu \partial_\rho u_\sigma$, as well as the vector/axial chemical potential $\mu_V \equiv (\mu_++\mu_-)/2$, $\mu_A \equiv (\mu_+-\mu_-)/2$, then the equilibrium hydrodynamic quantities are, to leading order in $\hbar$:
\begin{eqnarray}\begin{split}
&J_{\mathrm{eq},V}^{\mu}  \,=\,
 	n_V u^\mu + \frac{\hbar}{2} \frac{\partial n_A}{\partial \mu_V} \omega^\mu
	\,,\qquad&
&T_{\mathrm{eq}}^{\mu\nu} \,=\,
	\varepsilon u^\mu u^\nu + P(u^\mu u^\nu - g^{\mu\nu}) + \frac{\hbar \,n_A}{4} (8 \omega^\mu u^\nu + T \epsilon^{\mu\nu\sigma\lambda}\varpi_{\sigma\lambda}) \,,\\
& J_{\mathrm{eq},A}^{\mu} \,=\,
	n_A u^\mu + \frac{\hbar}{2} \frac{\partial n_A}{\partial \mu_A} \omega^\mu
	\,,\qquad&
&S_{\mathrm{eq}}^{\lambda\mu\nu} \,=\,
	 \frac{n_A}{2} \epsilon^{\lambda\mu\nu\sigma} u_\sigma 
	+ \frac{\hbar}{4} \frac{\partial n_A}{\partial \mu_A} \epsilon^{\lambda\mu\nu\sigma}\omega_\sigma\,.
\end{split}\end{eqnarray}
In above equations, the quantum corrections to the vector and axial currents are collectively known as the Chiral Vortical Effect, (see e.g.~\cite{Son:2009tf}). 
In particular, even in the purely neutral case $\mu_V = \mu_A = 0$, the quantum correction to the axial current, $\hbar( T^2 \omega^\mu/6)$, is non-vanishing.
Noting that this leads to non-zero spin density $u_\lambda S_{\mathrm{eq}}^{\lambda\mu\nu} = \hbar T^3 \varpi^{\mu\nu}/12$, such a quantum correction term induces the spin-vorticity alignment.
On top of existing chiral-hydro that includes anomalous transport terms in current and axial current, the above equations also include quantum corrections to the stress tensor, which is important for a self-consistent study of chiral effects.
Last but not the least, it is worth noting that these equations are causal and stable against linear perturbations~\cite{Shi.pre}:
this is non-trivial, as a first-order derivative term $\omega^\mu$ appears. Causality and stability of the equations follow from the fact that $\partial_\mu \omega^\mu =(1/2)\epsilon^{\mu\nu\rho\sigma}(\partial_\mu u_\nu) (\partial_\rho u_\sigma)$ does not contain second-order derivatives.

\section{Viscous Spin Hydrodynamics}\label{sec.viscous}
With ideal spin hydrodynamics obtained by taking equilibrium distributions, we further investigate the influence of non-equilibrium effect.
We start with taking the (14+6) moment formalism~\cite{Denicol:2010xn,Denicol:2012cn}, and expand the distribution function as %
\begin{eqnarray}
f^\pm &\equiv& f_\mathrm{eq}^\pm + f_\mathrm{eq}^\pm (1-f_\mathrm{eq}^\pm)
\bigg[ 
\mp	\hbar \lambda_{\Omega}^\pm \frac{\Omega^\pm \cdot p}{u\cdot p}
+	\lambda_{\nu}^\pm \nu_{\pm}^\mu p_\mu 
+	\lambda_{\pi}^\pm \pi^{\mu\nu} p_\mu p_\nu 
\bigg]\,,\label{eq.noneq}
\end{eqnarray}
where $\lambda_X$ are polynomials of co-moving energy $(u\cdot p)$ that are determined by matching the distribution function with dissipative quantities 
$\pi^{\mu\nu}$, $\nu^\mu$ and $\Omega^\mu$.
In particular, $\Omega^\mu$ characterizes the difference between spin polarization density and its thermal equilibrium value, while we expect no bulk pressure $\Pi$ when discussing massless particles.
Then we include collision terms in chiral kinetic equations
\begin{equation}
p^\mu \partial_\mu f_\pm
	\pm \hbar \frac{\epsilon^{\mu\nu\rho\sigma} p_\nu (\partial_\rho u_\sigma)}{4\, u \cdot p} \partial_\mu f_\pm
	 = \mathcal{C}_\pm[f_+, f_-] \,,
\end{equation}
and obtain the equations of motion for dissipative terms
\begin{eqnarray}
\Delta^{\mu\nu}_{\alpha\beta}\hat{\mathrm{d}} \pi^{\alpha\beta} + \tau_{\pi}^{-1} \pi^{\mu\nu} &=& \frac{2 \,\eta\, \sigma^{\mu\nu}}{\tau_{\pi}} + \text{[2nd order terms]} \,,\\
\Delta^{\mu}_\alpha \hat{\mathrm{d}} \nu_{\pm}^{\alpha} + \tau_{\nu,\pm}^{-1}  \nu_{\pm}^{\mu}  &=& \frac{\sigma}{\tau_{\nu,\pm}} \partial^\mu \frac{\mu_\pm}{T} +  M_{\nu,\pm}  \nu_{\mp}^{\mu}
+ c_{\Omega,\pm} \Omega_{+}^{\mu} + c_{\Omega,\pm} \Omega_{-}^{\mu} + \text{[2nd order terms]}  \,,\\
\Delta^{\mu}_\alpha \hat{\mathrm{d}} \Omega_{\pm}^{\alpha} + \tau_{\Omega,\pm}^{-1}  \Omega_{\pm}^{\mu}  &=&  M_{\Omega,\pm} \Omega_{\mp}^{\mu}   + \text{[2nd order terms]}  \,,
\end{eqnarray}
with coefficients $\tau_{\pi}$, $\tau_{\nu}$, $\tau_{\Omega}$, $\eta$, $\sigma$, $M_\nu$, $M_\Omega$, and $c_\Omega$ being functions of temperature $T$ and chemical potentials $\mu_\pm$~\cite{Shi.pre}.
Especially, we find that spin corrections to the EoM of the shear tensor $\pi^{\mu\nu}$ appear only up to second-order in gradient expansion, but contribute to first-order terms in the EoM of dissipative currents $\nu^\mu_\pm$.
Besides, the vector $\Omega^\mu$ relaxes to zero, which indicates that the full polarization tensor could return to the equilibrium limit.

Additionally, from the non-equilibrium distribution function (\ref{eq.noneq}), we obtain the RH/LH currents and energy-momentum tensor:
\begin{eqnarray}
J_{\pm}^{\mu} 
&=& n_\pm u^\mu + \nu_\pm^\mu 
	\pm \frac{\hbar}{2} \frac{\partial n_\pm}{\partial \mu_\pm} \omega^\mu
	\pm \frac{\hbar}{2} \epsilon^{\mu\rho\sigma\lambda} u_\rho \partial_\sigma \Big(\frac{G_{4,1}^{(1),\pm}}{D_{3,1}^\pm} \nu_{\pm,\lambda}\Big)
	\pm \frac{\hbar J_{2,2}^\pm}{4J_{4,2}^\pm} \Big( \epsilon^{\mu\rho\sigma\lambda} u_\rho {\sigma_{\sigma}}^{\xi}\pi_{\lambda\xi}
	- \pi^{\mu\lambda} \omega_{\lambda} \Big)\,,
\end{eqnarray}
\begin{eqnarray}
T^{\mu\nu} 
&=&	\epsilon u^\mu u^\nu + P (u^\mu u^\nu - g^{\mu\nu}) + \pi^{\mu\nu} 
	+\frac{\hbar \,n_A}{4} (8 \omega^\mu u^\nu + T \epsilon^{\mu\nu\sigma\lambda}\varpi_{\sigma\lambda})
	+ \frac{4\hbar}{5}\omega^\mu (\nu_+^{\nu} - \nu_-^{\nu}) 
\nonumber\\&&
	+ \hbar (u^\mu \Omega_+^\nu + u^\nu \Omega_+^\mu)
	- \hbar (u^\mu \Omega_-^\nu + u^\nu \Omega_-^\mu)
	+ \frac{\hbar}{2} \epsilon^{\mu\rho\sigma\lambda} u_\rho \Delta^{\nu\xi}\partial_\sigma\bigg[
	 \Big(\frac{J_{3,2}^+}{2J_{4,2}^+} -\frac{J_{3,2}^-}{2J_{4,2}^-} \Big)\pi_{\lambda\xi}  \bigg]
\nonumber\\&&
	- \frac{\hbar}{10} \epsilon^{\mu\nu\rho\sigma} u_\rho (\partial_\sigma u^\lambda) 
	(\nu^+_{\lambda} - \nu^-_{\lambda})
	+ \frac{2\hbar}{5} \epsilon^{\mu\lambda\rho\sigma} u_\rho (\partial_\sigma u^\nu) (\nu^+_{\lambda} - \nu^-_{\lambda})
	+  \frac{\hbar}{2} \epsilon^{\mu\rho\sigma\lambda} u_\rho u^\nu \partial_\sigma( \nu^+_{\lambda} -  \nu^-_{\lambda})
	 \,.
\end{eqnarray}
In the above expressions, $J^\pm_{n,q}$, $D^\pm_{n,q}$, and $G^{(q),\pm}_{n,m}$ denote for thermal integrals, and are functions of temperature $T$ and chemical potentials $\mu^\pm$~\cite{Shi.pre,Denicol:2012cn}.

\section{Summary}
In this work, we start from a 14+6 moment expansion formalism and obtain the second-order viscous spin hydrodynamics from a system of massless Dirac spinors.
In such a system, the spin alignment effect could be treated in the same framework as for chiral hydrodynamics, but with non-trivial quantum corrections to the stress tensor.
With a framework that self-consistently solves the evolution of systems containing spin degrees of freedom, we expect future work to precisely quantify both global and local polarization rates of final-state hadrons created in heavy-ion collisions.

\vspace{0.1in}
\textbf{Acknowledgments.}
This work is supported by the Natural Sciences and Engineering Research Council of Canada.
SS would like to thank Wojciech Florkowski, Xu-Guang Huang, Jinfeng Liao, Jorge Noronha, Qun Wang, and Yi Yin for helpful discussions.




\end{document}